\documentclass[%
 reprint,
 amsmath,amssymb,
 aps,
 pra,
]{revtex4-2}

\usepackage{graphicx,subcaption}
\usepackage{dcolumn}
\usepackage{bm}

\usepackage{mathtools}

\usepackage{braket}
\usepackage{algpseudocode}

\usepackage[normalem]{ulem} 
\usepackage{xcolor}

\begin{document}

\preprint{APS/123-QED}

\title{Multiple-time Quantum Imaginary Time Evolution}
\thanks{A footnote to the article title}%

\author{Julio Del Castillo}
\affiliation{
Universidad Nacional de Educación a Distancia, Spain
\\ \texttt{jdelcasti26@alumno.uned.es}}

\author{Mats Granath}
\affiliation{
Department of Physics, University of Gothenburg, Sweden
}

\author{Evert van Nieuwenburg}
\affiliation{
 $\langle aQa^L \rangle$ at Leiden Institute of Physics and the Leiden Institute of Advanced Computer Science, Leiden University, The Netherlands
}

\date{\today}

\begin{abstract}
Quantum Imaginary-Time Evolution (QITE) is a powerful method for preparing ground states on quantum hardware. However, executing QITE has costly measurement budgets for general Hamiltonians. Both fidelity and computational cost are strongly dependent on the definition of suitable local domains and Hamiltonian partitions. In this work, we introduce the Multiple-Time QITE algorithm (MT-QITE). We show how using more than one imaginary time substantially improves the fidelity of the resulting ground state as well as the measurement overhead with respect to the previously published QITE algorithm, while preserving its deterministic character and its independence from \textit{ad hoc} ansätze. Moreover, unlike QITE and other QITE-based algorithms, MT-QITE is parallelizable, and we show that even in Hamiltonians with non-local interactions, partitioning may entail a computational advantage.
\end{abstract}



\maketitle


\section{\label{sec:intro}Introduction}


Quantum simulation stands as one of the most promising applications of quantum computers to tackle intractable problems of commercial and industrial relevance to date \cite{alexeev_2024, mcardle_2020}. The preparation of ground states of physical systems is a task of paramount importance in quantum simulation, as many physical and chemical properties of the system are derived from its lowest energy and the associated wavefunction. A wealth of numerical methods has been developed to obtain ground states since the early days of quantum mechanics \cite{martin_2020}. However, they either tend to incur exponential memory and runtime overheads with system size, or they fail to describe certain systems where correlations are not captured correctly with independent-particle approximations.
Several quantum algorithms have been proposed to obtain ground states both for the coming fault-tolerant quantum computers \cite{campbell_2017} and the current noisy intermediate-scale devices (NISQ) \cite{Preskill_2018}. Among the former Quantum Phase Estimation \cite{nielsen_2010} is set to become a pivotal technique once full-fledged quantum devices are available, among the latter the Variational Quantum Eigensolver (VQE) \cite{peruzzo_2014} and Imaginary Time Evolution (ITE) related quantum algorithms \cite{motta_2020, liu_2021, mcardle_2019} have attracted a lot of attention, as they are suited for current error-prone hardware with a limited budget in circuit depth. Even in the fault-tolerant era, the finite execution times of error-corrected one- and two-qubit gates will limit algorithm performance, and notably measurement operations will impose further latency. The fewer quantum gates and the smaller the measurement budget, the faster a routine can be executed. Moreover, the performance of any of the ground state algorithms mentioned above depends crucially on the overlap of the initial state with the target ground state. Thus, the ability to prepare quantum states with a given fidelity and an optimal gate count is instrumental in the digital quantum simulation of any physical system that requires ground state preparation.
The first implementation of ITE on quantum hardware was Variational Quantum Imaginary Time Evolution (VQITE), proposed by McArdle et al. \cite{mcardle_2019} in 2019. The protocol involves mimicking ITE on an ansatz subspace following an equation of motion based on the McLachlan variational principle \cite{mclachlan_1964}, whose parameters are determined by measurements on a quantum register and processed classically. Several more recent works have developed and successfully demonstrated the protocol \cite{gacon_2023, gomes_2021, selvarajan_2021, yuan_2019}. Motta et al. introduced the QITE algorithm \cite{motta_2020} in 2020, also a hybrid classical-quantum protocol, but ansatz-free unlike VQITE. QITE approximates the non-unitary imaginary-time evolution by a real-time evolution (RTE) defined by a linear combination of Pauli operators whose parameters are also determined by tomography and processed classically. Moreover, the QITE subroutine may be used for the determination of excited states with the QLanczos algorithm and thermal states with QMETTS, both introduced in \cite{motta_2020}. Many refinements and applications of QITE have appeared in the literature \cite{benedetti_2021, cao_2022, kamakari_2022, gomes_2020, kolotouros_2025, nishi_2021, sun_2021, yeter_2022}. Other approaches to ITE on quantum computers include the following: (1) Probabilistic Imaginary Time Evolution (PITE) \cite{liu_2021, kosugi_2023, leadbeater_2024}, where an ancillary qubit is employed and the probabilistic outcome of its measurement determines the collapse of the rest of the quantum register to the ITE-evolved state with a desired accuracy and success rate (2) Adiabatic QITE \cite{hejazi_2024}, where an auxiliary Hamiltonian is built, whose ground state is the ITE-evolved state of the target Hamiltonian, and classical numerical integration of its equation of motion provides the parameters to implement RTE on quantum hardware to a desired precision (3) Double-Bracket QITE (DB-QITE) \cite{gluza_2024} proposes an ansatz-free, tomography-free protocol to implement QITE at the expense of introducing an exponentially large number of gates with the number of steps (4) and more recently Constant-Depth QITE using dynamic fan-out circuits \cite{lund_2026}, especially well-suited for diagonal Hamiltonians.

In this work, we introduce the Multiple-Time QITE (MT-QITE) algorithm. Our heuristic proposal is based on the previously reported QITE method as defined in \cite{motta_2020}, allowing several imaginary times in the trotterization of the problem Hamiltonian in a variational fashion. These new degrees of freedom allow attaining a better upper bound for the ground energy when dealing with non-commuting Hamiltonian partitions and, in addition, result in a significant reduction of the computational cost required to achieve maximum fidelity. Moreover, MT-QITE uses significantly fewer reference states on which measurements are performed, allowing savings in the measurement budget and the parallelization of the routine. Our proposal is suited for both NISQ and fault-tolerant devices and, in the same conditions as the original QITE, remains deterministic while improving fidelity and reducing the measurement count budget. Unlike other variational approaches such as VQE, MT-QITE does not require the \textit{ad hoc} selection of an ansatz, it is optimally determined at each step. The manuscript is organized as follows: section \ref{sec:theo} reviews the principles of the QITE algorithm and presents the MT-QITE algorithm for general Hamiltonians and quantum chemical ones. Section \ref{sec:results} presents simulation results on 4, 6 and 8-qubit physical systems. Section \ref{sec:conclusion} discusses the outlook. Supplementary material in section \ref{sec:supplement} is provided with an alternative derivation of the QITE formulas that involve exclusively measurements of hermitian operators and comparison to related work.

\begin{figure*}[t]
    \centering
    \includegraphics[width=0.7\textwidth]{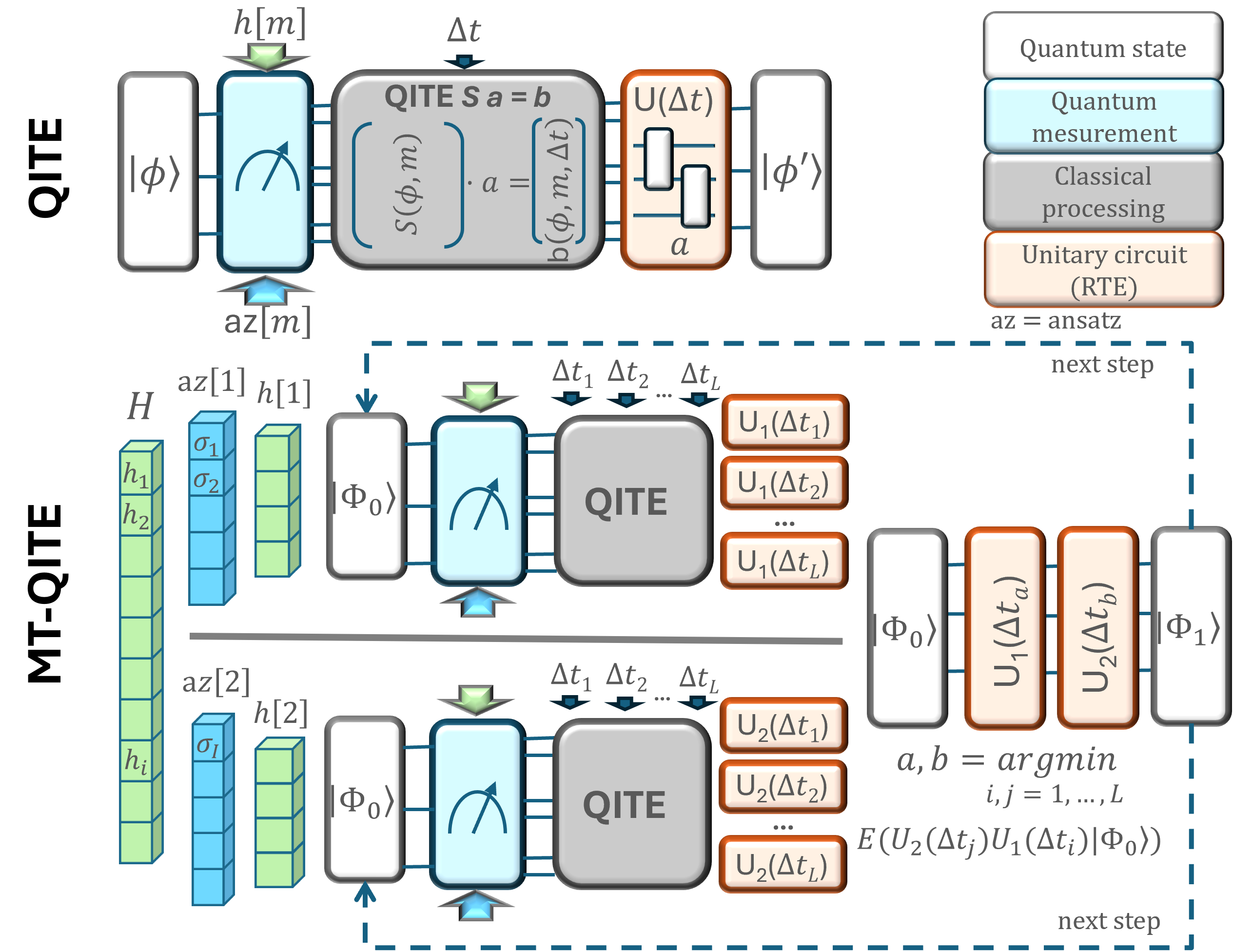}
    \caption{\label{fig:parallelmtq}Illustration of the MT-QITE algorithm and its relation to the QITE routine. Given a quantum state, a Hamiltonian term $h[m]$ and a basis ansatz $az[m]$, QITE produces an evolved state through a real-time evolution (RTE) whose parameters are determined by tomography, the desired time step size $\Delta t$, and the classical resolution of a linear system of equations. The idea behind MT-QITE is to use the same reference quantum state for all terms in the Hamiltonian partition (only two are depicted for clarity), and to obtain the RTE parameters for a set of different time step sizes, as unlike QITE, MT-QITE only adds classical processing for a new time step size. The evolved state is the one on which the expectation value of the Hamiltonian is minimum, possibly using a different time step size per partition term.}
\end{figure*}

\section{\label{sec:theo}Theory}

In this section, the theoretical principles underlying QITE-based algorithms are presented. In subsection \ref{subsec:qite} the general Imaginary Time Evolution (ITE) method is reviewed, along with previously reported primitives for its implementation on quantum hardware with the QITE algorithm \cite{motta_2020, sun_2021}. In subsection \ref{subsec:mtqite} the Multiple-Time QITE algorithm (MT-QITE) is introduced, based on the QITE algorithm as a sub-routine. Subsection \ref{subsec:mtqitechem} presents a general reformulation of the QITE equations that is especially well-suited for quantum chemistry Hamiltonians and coupled-cluster-based ansätze.

\begin{figure*}[t]
  \centering
  \begin{subfigure}{0.32\textwidth}
    \includegraphics[width=\linewidth]{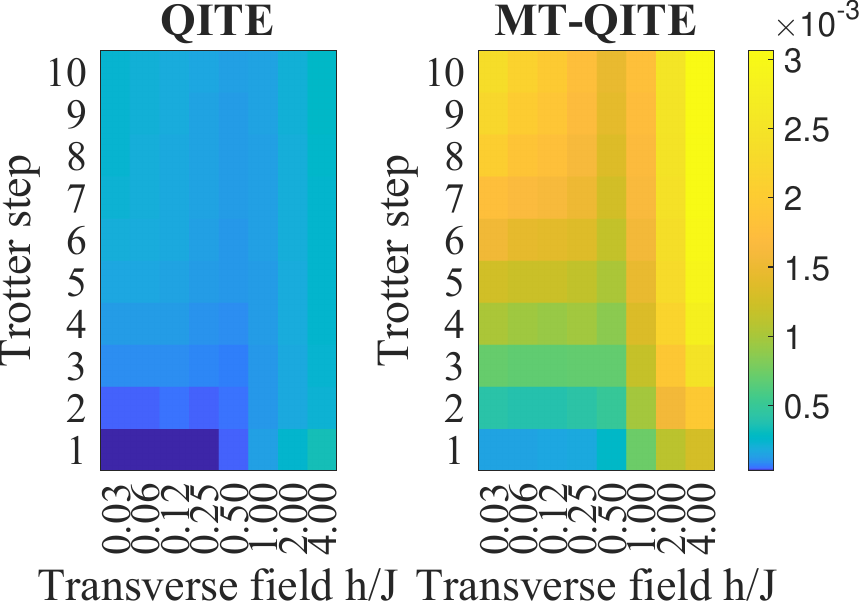}
    \caption{Transverse-field Ising model}
    \label{fig:heatma}
  \end{subfigure}\hfill
  \begin{subfigure}{0.32\textwidth}
    \includegraphics[width=\linewidth]{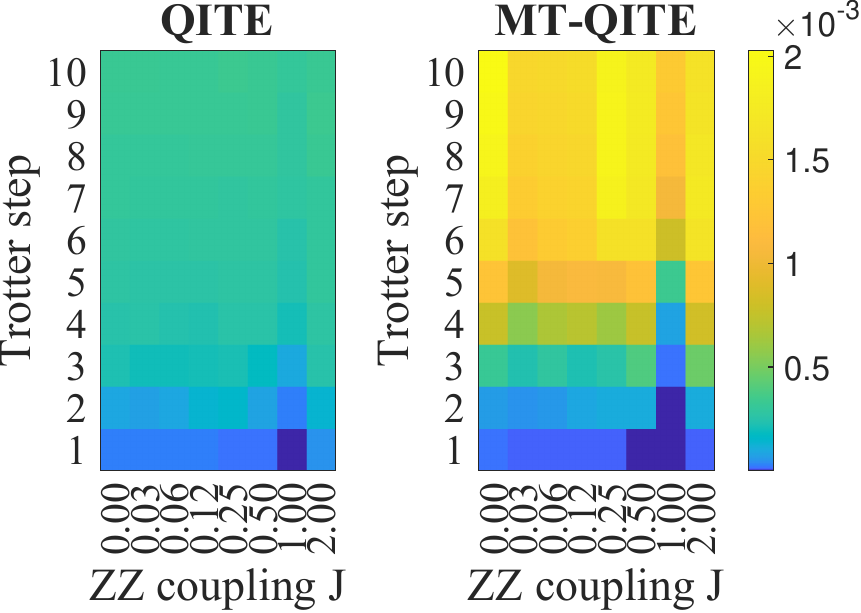}
    \caption{XXZ Heisenberg model}
    \label{fig:heatmb}
  \end{subfigure}\hfill
  \begin{subfigure}{0.32\textwidth}
    \includegraphics[width=\linewidth]{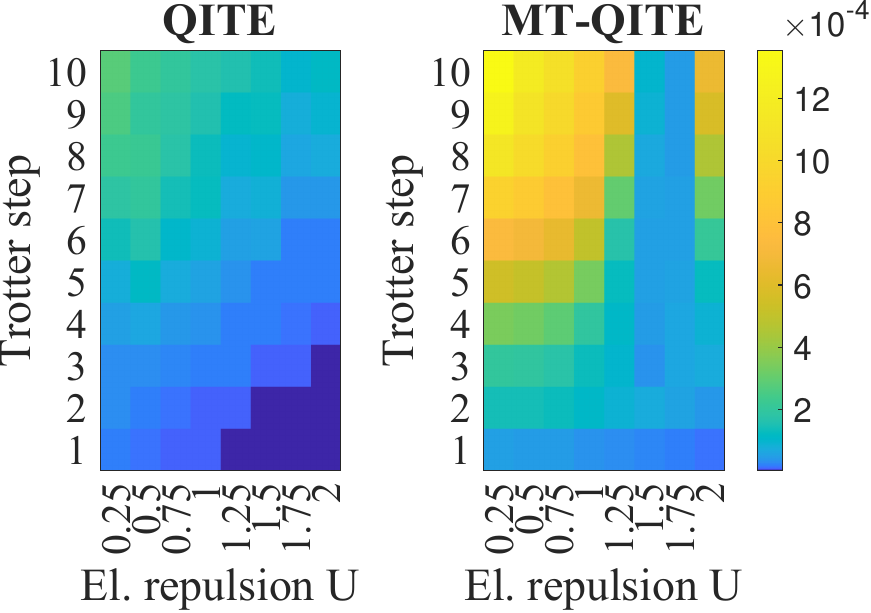}
    \caption{Fermionic Hubbard model}
    \label{fig:heatmc}
  \end{subfigure}

 \caption{Fidelity per extra Pauli measurement per Trotter step in logarithmic scale for a selection of models and parameter regimes, namely: (a) Transverse field for the Ising model h/J, (b) ZZ coupling for the XXZ Heisenberg model J, (c) Electrostatic repulsion for the fermionic Hubbard model U. For the Ising and Heisenberg models simulations become more challenging around the phase transition (h/J = 1 and J= 1 respectively), and for the Hubbard model when strong repulsion is present. Improvement in fidelity per extra measurement is clearly superior for MT-QITE.}

  \label{fig:fig_heatm}
\end{figure*}

\subsection{QITE}
\label{subsec:qite}

The classical ITE algorithm is a tried-and-tested technique for obtaining the ground state of a quantum system as the long-time evolution of the imaginary-time Schrödinger equation:
  
\begin{equation}
-\partial_\beta\ket{\Phi(\beta)} = \hat{H}\ket{\Phi(\beta)}
\label{eq:sch_imtime}
\end{equation}

\noindent 
whose normalized solution can be expressed as the long-time limit of the imaginary-time evolution operator $e^{-\beta\hat{H}}$ applied to an inital state $\ket{\Phi(0)}$ as:
  
\begin{equation}
\ket{\Psi} = \lim\limits_{\beta \to \infty}\frac{e^{-\beta\hat{H}}\ket{\Phi(0)}}{{\lVert}e^{-\beta\hat{H}}\ket{\Phi(0)}\rVert}
\label{eq:limit_imtime}
\end{equation}

\noindent 
provided there is an overlap between the initial state and the ground state. 

\begin{equation}\label{eq:overlap}  
\braket{\Psi|\Phi(0)}\neq{0}
\end{equation}

Classically, the method is numerically stable and guaranteed to converge, but its resource requirements scale exponentially with system size due to the exponential growth of the state-vector dimension. The main difficulty for the implementation of ITE on a quantum computer is that the evolution operator is not unitary, while by construction, all operations available on quantum hardware, except for measurement, are unitary. To overcome this constraint, the QITE algorithm \cite{motta_2020} performs a Suzuky-Trotter decomposition of the Hamiltonian, where a unitary operator is optimally calculated at each step to follow the trajectory of the non-unitary imaginary-time evolution of the state vector.
  
Consider a Hamiltonian that can be expressed as a sum of local terms acting on at most $k$ qubits:
  
\begin{equation}
\hat{H} = \sum_m{\hat{h}[m]}
\label{eq:local_ham}
\end{equation}

\noindent
where each $\hat{h}[m]$ may include several Pauli strings. The imaginary-time evolution operator can then be expressed as a Trotter decomposition:
  
\begin{equation}
e^{-\beta\hat{H}} = (e^{-\Delta\tau{\hat{h}[1]}}e^{-\Delta\tau{\hat{h}[2]}}...)^n + \mathcal{O}(\Delta\tau^2); n = \frac{\beta}{\Delta\tau}
\label{eq:trotter}
\end{equation}

\noindent
The cornerstone of QITE lies on the approximation of the normalized state, after a single step of the imaginary-time evolution operator, by the application of a unitary operator acting on a neighbourhood of the $k$ qubits involved in $\hat{h}[m]$, with $D$ qubits in total i.e.
  
\begin{equation}
\ket{\Phi'} = \frac{e^{-\Delta\tau{\hat{h}[m]}}\ket{\Phi}}{\lVert e^{-\Delta\tau{\hat{h}[m]}}\ket{\Phi}\rVert} \approx e^{-i\Delta\tau{\hat{A}[m]}}\ket{\Phi}
\label{eq:unit_approx}
\end{equation}

\noindent
$\hat{A}[m]$ can then be expanded in terms of the Pauli basis $\{\hat{\sigma_i}\}$, spanning $4^D$ Pauli strings:
  
\begin{equation}
\hat{A}[m] = \sum_{i_1...i_D}a[m]_{i_1...i_D}\hat{\sigma}_{i_{1}}...\hat{\sigma}_{i_{D}} = \sum_I{a[m]}_I\hat{\sigma}_I
\label{eq:pauli_expansion}
\end{equation} 

\noindent
Expanding the exponential operators to order 1 in ~(\ref{eq:unit_approx}) and minimizing the difference norm of the two states, coefficients $a[m]_{i_1...i_D}$ can be derived classically by solving a linear system of equations that involves measurements on $\ket{\Phi}$:

\begin{eqnarray}
\textbf{S}\textbf{a}[m]=\textbf{b}
\\
c = 1 - 2\Delta\tau\bra{\Phi}\hat{h}[m]\ket{\Phi}
\\
S_{I,J} = 2 Re \bra{\Phi}\sigma^{\dagger}_I\sigma_J\ket{\Phi}
\\
b_{I} = 2 Im \frac{1}{\sqrt{c}} \bra{\Phi}\sigma^{\dagger}_I\hat{h}[m]\ket{\Phi}
\label{eq:linear_sys1}
\end{eqnarray}

It is worth noting that the number of unique expectation values that need to be measured at each step is $O(4^{D})$. This overhead can be optimized taking into account symmetries in the problem Hamiltonian, as it will be discussed below. There are two sources of error in the algorithm, the Trotter error and the accuracy of the unitary approximation over the domain $D\leq N$, $N$ being the total number of qubits. It follows from Uhlmann's theorem \cite{uhlmann_1976} that the approximation holds if $D$ is sufficiently large, becoming exact if $D = N$.
  
Both accuracy and numerical stability can be improved considerably by expanding the exponentials up to order 2 on the left side of ~(\ref{eq:unit_approx}) introducing some extra measurement overhead that grows quadratically with the number of Pauli strings in the Hamiltonian term \cite{sun_2021}:

\begin{eqnarray}
c = 1 - 2\Delta\tau\bra{\Phi}\hat{h}[m]\ket{\Phi} + 2\Delta\tau^2\bra{\Phi}\hat{h}[m]^2\ket{\Phi}
\label{eq:linear_system2b}
\\
S_{I,J} = 2 Re \bra{\Phi}\sigma^{\dagger}_I\sigma_J\ket{\Phi}
\label{eq:linear_system3b}
\\
b_{I} = 2 Im \frac{1}{\sqrt{c}} \bra{\Phi}{\sigma^{\dagger}_I\hat{h}[m] - \frac{\Delta\tau}{2}\sigma^{\dagger}_I\hat{h}[m]^2}\ket{\Phi}
\label{eq:linear_sys2}
\end{eqnarray}  

\noindent
Note that exact order 2 on both sides of ~(\ref{eq:unit_approx}) would not lead to a linear system of equations but to a cubic one. In this work, the second-order formulas above are privileged, as the numerical instability introduced by the order-one implementation poses greater problems in the simulation of some of the analyzed physical models. The quadratic overhead introduced by $\hat{h}[m]^2$ can be kept in check by choosing a suitable partition for the Hamiltonian.

It is important to notice that the derivation of the equations above involves taking expected values of non-hermitian operators. This is mathematically consistent in the context of normal bounded linear operators in a Hilbert space. An operator $\hat{N}$ is said to be normal if it commutes with its adjoint, $[\hat{N}, \hat{N}^{\dagger}] = 0$, and it is diagonalizable, possibly with complex eigenvalues \cite{alt_2016}.

In general, the linear system defined by the above equations has a non-zero null space, and it must be solved using an appropriate method such as the Moore-Penrose pseudo-inverse algorithm \cite{penrose_1955}. In many cases, the resulting circuit depth can be reduced by identifying solutions that encompass more zero values, at the expense of introducing numerical instability in some problems and parameter regimes. In this work, we prioritize low-depth as much as possible.

The measurement overhead of the QITE algorithm can be greatly reduced exploiting $\mathbb{Z}_{2}$ symmetries in the Hamiltonian, as reported in \cite{sun_2021}. The underlying principles are based on the stabilizer formalism \cite{nielsen_2010} and it is possible to prove that it is not necessary to consider a $4^D$ x $4^D$ linear system involving the $4^D$ Pauli strings, but only a subset of them obtained as follows: Out of all the possible Pauli strings identify those that commute with all the terms in the Hamiltonian and among themselves, this forms the stabilizer $S$, which is an abelian subgroup of the Pauli group. Then check which strings commute with the stabilizer operators, this defines the normalizer $N(S)$, which is a subgroup of the Pauli group. Form the quotient group $N(S)/S$ and select only one Pauli string from each coset. Pick an initial state to start the QITE algorithm from the invariant subspace, defined by the stabilizer (the eigenspace with well defined $+1/-1$ eigenvalues for each operator in $S$).

\subsection{MT-QITE}
\label{subsec:mtqite}

Recalling ~(\ref{eq:limit_imtime}) and ~(\ref{eq:trotter}), and ignoring the normalizing factors, the ground state of a given system can be approximated as:
  
\begin{equation}
\ket{\Psi} \approx (e^{-\frac{\beta}{n}{\hat{h}[1]}}e^{-\frac{\beta}{n}{\hat{h}[2]}}...)^n\ket{\Phi(0)}
\label{eq:ftqite1}
\end{equation}

\noindent
with the error scaling as $O(\frac{\beta^2}{n})$. In standard QITE, as published in \cite{motta_2020}, the unitary operator at each step is determined by tomography on the so far evolved state, i.e.

\begin{equation}
\ket{\Psi''} \propto e^{-\frac{\beta}{n}{\hat{h}[m]}}\ket{\Psi'}
\label{eq:ftqite2}
\end{equation}

\noindent
and the number of gates in the obtained circuit grows linearly with each step.

In practice, one fixes a maximum $n$ and chooses a small $\Delta\tau = \beta/n$. Measuring the Hamiltonian on the resulting states, it becomes clear that there is always an optimum time step size that has to be determined by running the algorithm for several values of  $\Delta\tau$. So, to some extent, given $n$,  finding a good estimate of the ground with QITE is a variational problem.

The heuristic algorithm we propose is based on the following strategy: suppose that we have a suitable initial state $\ket{\Phi(0)}$ with a convenient overlap with the ground state and that we can define a suitable partition of the Hamiltonian $\hat{H} = \hat{h}[1] + \hat{h}[2]$. This partition may either minimize the norm of the commutator $[\hat{h}[1], \hat{h}[2]]$ and thus the Trotter error, or act on well-defined qubit domains, so that the number of measurements is reduced (e.g. organising two partitions that require two reduced sets of Pauli operators by a factor of 2 results in two linear systems involving a quarter of measurements). In what follows, we consider only a two-term partition and a single Trotter step for clarity.

QITE's ansazt is related, via unitary implementations of ITE, to the following wavefunction:

\begin{equation}
\ket{\Psi(t)} = \frac{e^{-t\hat{h}[2]}e^{-t\hat{h}[1]}\ket{\Phi(0)}}{{\lVert}e^{-t\hat{h}[2]}e^{-t\hat{h}[1]}\ket{\Phi(0)}\rVert}
\label{eq:qite_ansatz}
\end{equation}

and it will produce an upper bound to the ground energy:

\begin{equation}
\min_t \bra{\Psi(t)}H\ket{\Psi(t)} 
\label{eq:qite_upbound}
\end{equation}

\noindent
It is natural to assume that adding an additional degree of freedom to time, we can obtain a smaller estimate of the ground energy, as the variational ansatz is more flexible, i.e.

\begin{equation}
\ket{\Psi'(t_1,t_2)} = \frac{e^{-t_2\hat{h}[2]}e^{-t_1\hat{h}[1]}\ket{\Phi(0)}}{{\lVert}e^{-t_2\hat{h}[2]}e^{-t_1\hat{h}[1]}\ket{\Phi(0)}\rVert}
\label{eq:mtqite_ansatz}
\end{equation}

\noindent
and it will produce an upper bound to the ground energy:

\begin{equation}
\min_{t_1, t_2} \bra{\Psi'(t_1, t_2)}H\ket{\Psi'(t_1, t_2)} 
\label{eq:mtqite_upbound}
\end{equation}

\noindent
It is clear that the multiple-time ansatz includes QITE's along $t_1=t_2$, and its upper bound for the energy will be less than or equal to QITE's. This is the cornerstone of the MT-QITE algorithm described in what follows.

In QITE, the quantum state used to determine the linear system is updated at each step as illustrated by ~(\ref{eq:ftqite2}). In MT-QITE, that state update is not necessarily advantageous to the resulting fidelity. Alternatively, using the same fixed reference state for the whole Trotter step introduces a substantial computational advantage. See FIG. \ref{fig:parallelmtq} and FIG. \ref{fig:alg}. On the one hand, given a Hamiltonian term, its associated domain definition and its Pauli basis, the same set of measurements can be used for any time step size. On the other, many of the Pauli strings needed to compute the linear system coefficients appear several times in a given Trotter step, for different terms of the Hamiltonian, and may be reused. This is only possible because the reference state is shared as illustrated in FIG. \ref{fig:parallelmtq} and FIG. \ref{fig:alg}.

In practice, MT-QITE amounts to running QITE on each partiton term $m$, agnostic of the others, using the same reference state, determining the real-time evolution (RTE) unitaries, and obtaining one time step size per term that minimizes:
\begin{eqnarray}
\min_{t_1, t_2} \bra{\Phi(0)}e^{+i t_2\hat{A}[2]}e^{+i  t_1\hat{A}[1]}{\hat{H}}e^{-i t_1\hat{A}[1]}e^{-i t_2\hat{A}[2]}\ket{\Phi(0)}
\label{eq:ftqite7a}
\end{eqnarray} 

Having the same reference state for all the Hamiltonian terms in a Trotter step makes MT-QITE parallelizable, and in addition, it allows an additional reduction of the computational cost by symmetry. For example, consider the Heisenberg model on a $1D$ chain with open boundaries. Clearly, this system has site inversion symmetry, i.e. the Hamiltonian commutes with the site inversion operator defined as $R\ket{q_N ... q_2 q_1} = \ket{q_1 ... q_{N-1} q_N}$. The eigenvalues of $R$ are $+1$ or $-1$. Choosing a well-defined eigenvalue for the initial state, it is by construction a constant of motion, as $[H, R] = 0$. In this case, MT-QITE can be run without measuring those terms in the partition that are the $R$ symmetry of a previously measured one. The reference state being the same in MT-QITE for every term, application of the $R$ symmetry to the previously obtained RTE operators produces the same result as the measurement and linear system protocol. This cannot be applied to standard QITE, as the reference state is updated at each step.

Given a partitioned Hamiltonian $\hat{H}=\hat{h}[1]+\hat{h}[2]$, an initial state $\ket{\Phi(0)}$ and a set of L differnt time step sizes $\{\Delta{t_l}:l=1,\ldots,L\}$, the algorithm we propose can be implemented as described in FIG~\ref{alg:mtqite}:

\begin{figure}[t]
\caption{MT-QITE Algorithm} 
\label{alg:mtqite}
\begin{algorithmic}[1]
\Require{$\ket{\Phi(0)}, \hat{h}[1], \hat{h}[2], \{\Delta{t_l}:l=1,\ldots,L\}$}
\State{$\ket{\Psi}\leftarrow\ket{\Phi(0)}$}
\State{// For all Trotter steps}
\For{$n = 1,2,\ldots,N$}
\State{$\ket{\Phi}\leftarrow\ket{\Psi}$}
\State{// For all terms in Hamiltonian partition}
\For{$m=1,2$}
\State{// For all time step sizes}
\For{$l=1,2,\ldots,L$}
\State{// Solve linear system}
\State{// (Measurements necessary for l = 1 only)}
\State{QITE($\ket{\Phi}, \hat{h}[m], \Delta{t_l}$)}
\State{// Save unitary RTE parameters}
\State{$U_{ml}\leftarrow$ Linear system solutions}
\EndFor
\EndFor
\State{// For all terms and time step sizes}
\For {$l_1=1,2,\ldots,L$}
\For {$l_2=1,2,\ldots,L$}
\State{Prepare $\ket{\Psi_{l_{1}l_{2}}} = U_{1l_{1}}U_{2l_{2}}\ket{\Phi}$}
\State{Measure $ E_{l_{1}l_{2}} = \bra{\Psi_{l_{1}l_{2}}}({\hat{h}[1]}+{\hat{h}[2]})\ket{\Psi_{l_{1}l_{2}}}$}
\EndFor
\EndFor
\State{$o_1,o_2 = argmin(E_{l_{1}l_{2}})$}
\State{$\ket{\Psi}\leftarrow U_{1o_{1}}U_{2o_{2}}\ket{\Phi}$}
\State{// Go to next Trotter step}
\EndFor
\end{algorithmic} 
\end{figure}

It is worth noting that the MT-QITE algorithm can also be applied to the trivial partition $\{\hat{H}, \emptyset\}$. In this case, the Trotter error is reduced at the expense of requiring larger linear systems and more measurements. This is true for both the original QITE algorithm and MT-QITE, but MT-QITE still benefits from a reduced measurement budget as a unique optimized wavefunction is used as input for every time step size in the following Trotter step.  

\begin{figure*}
  \centering
  \begin{subfigure}{0.48\textwidth}
    \centering
    \includegraphics[width=0.49\linewidth]{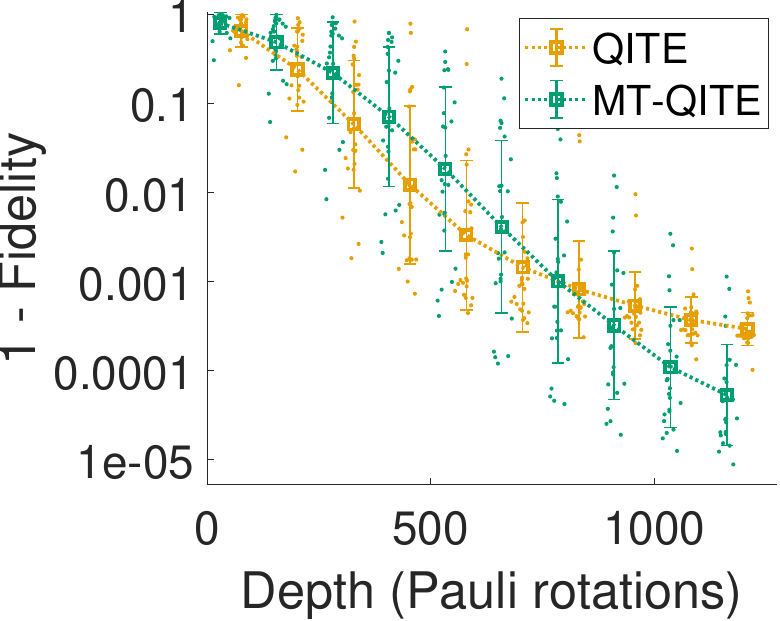}
    \includegraphics[width=0.49\linewidth]{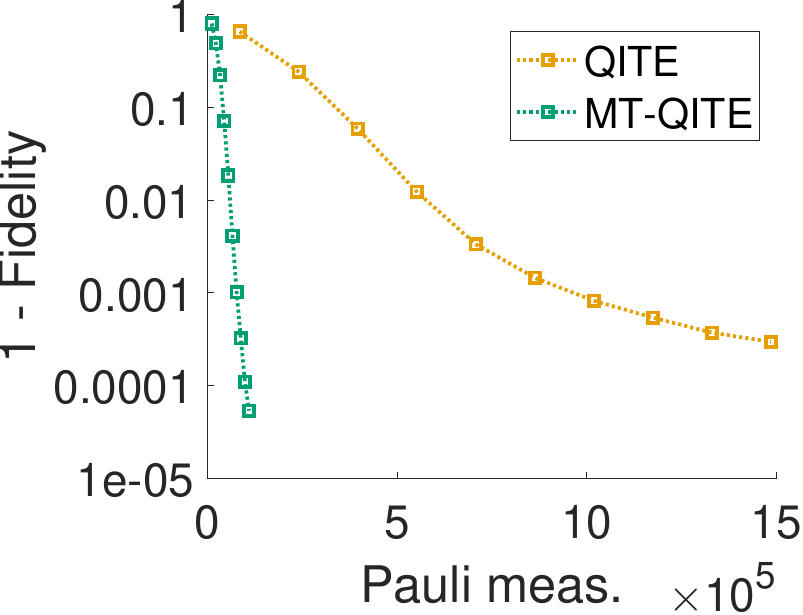}
    \caption{Heisenberg model}
    \label{fig:heis}
  \end{subfigure}\hfill
  \begin{subfigure}{0.48\textwidth}
    \centering
    \includegraphics[width=0.49\linewidth]{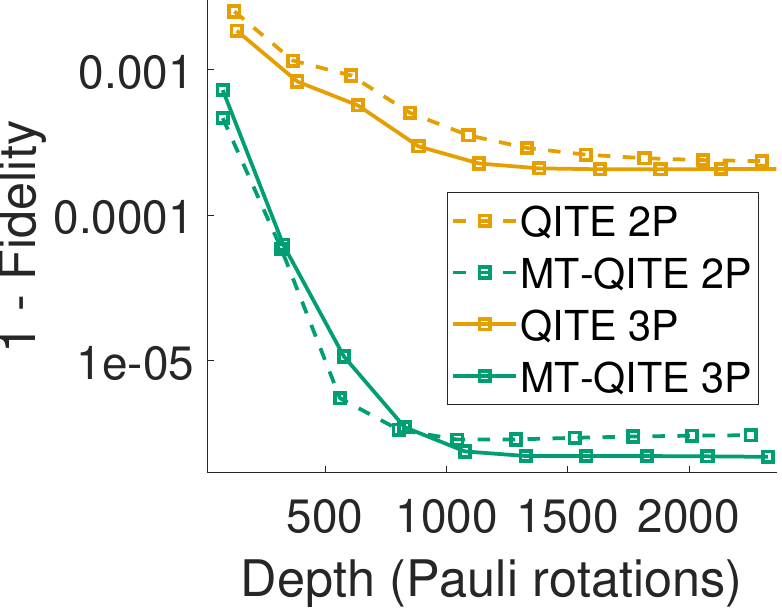}
    \includegraphics[width=0.49\linewidth]{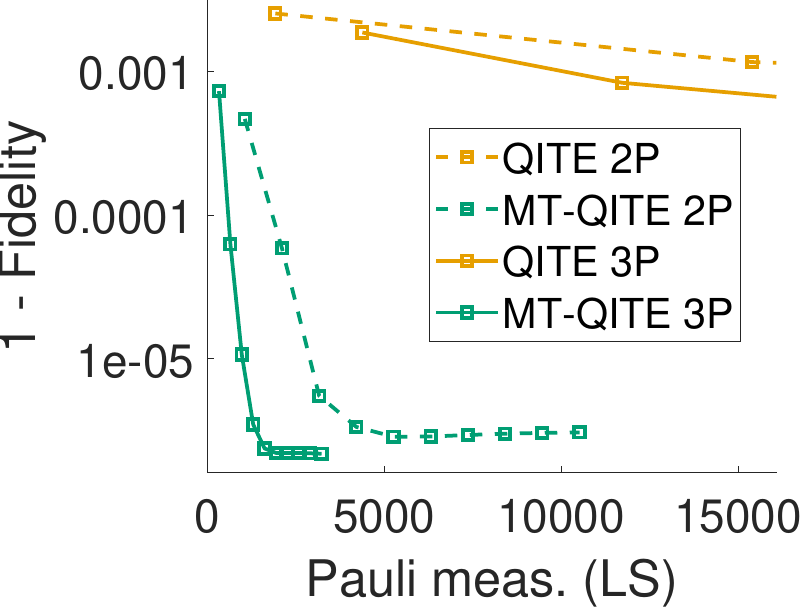}
    \caption{Ising model}
    \label{fig:ising}
  \end{subfigure}
  \caption{Ground state fidelity as a function of circuit depth expressed in no. of Pauli string rotations and cost in Pauli measurements (Necessary shots per Pauli not included). The MT-QITE algorithm (with native time-step optimization) is benchmarked with the previously reported QITE algorithm with optimal time-step at each Trotter step. (a) XXZ Heisenberg model (8 spins). A batch of 25 common initial states compatible with system symmetry is used, and 10 Trotter steps are run using a Hamiltonian partition with two terms. An improvement in fidelity of one order of magnitude is achieved with MT-QITE along with a reduction of the computational cost by a factor 10. MT-QITE outperforms QITE both in average and in the best sample observed. (b) Transverse-field Ising model (6 spins), using a single initial state and varying the partition configuration (two terms in 2P and three in 3P). Only linear system measurements are plotted for comparability, QITE lines display only 3 Trotter steps. In this case MT-QITE improves fidelity in more than two orders of magnitude and the measurement budget is considerably reduced by site inversion symmetry.}
   \label{fig:fig_main}
\end{figure*}

\subsection{MT-QITE for quantum chemistry}
\label{subsec:mtqitechem}

MT-QITE can be applied not only to general Hamiltonians with local interactions over a given domain, but also quantum chemical Hamiltonians. In this section we provide a general reformulation of QITE primitives in terms of anti-hermitian operators that is very well suited for both QITE and MT-QITE. 

The general form of the electronic Hamiltonian in quantum chemistry using chemist's notation \cite{mcardle_2020} is:

\begin{equation}
\hat{H}
=
\sum_{pq} h_{pq}\, \hat{a}_p^\dagger \hat{a}_q
+
\frac{1}{2} \sum_{pqrs} (pq|rs)\, \hat{a}_p^\dagger \hat{a}_q^\dagger \hat{a}_s \hat{a}_r,
\label{eq:2ndq_ham}
\end{equation}

\noindent
where $\hat{a}^\dagger$ and $\hat{a}$ are the fermionic creation and annihilation operators respectively, $h_{pq}$ and $(pq|rs)$ are the one- and two-electron integrals, and indices $p,q,r,s$ span a given basis set consisting of $N$ spin-orbitals.

QITE, like the Variational Quantum Eigensolver (VQE), has been successfully applied to quantum chemistry Hamiltonians using the Unitary Coupled-Cluster with Single and Double excitations (UCCSD) ansatz \cite{gomes_2020}. In this case, given $N$ spin-orbitals (qubits) the bottleneck of the algorithm is the number of terms in the Hamiltonian and in the ansatz, and both scale polynomially as $N^4$. The idea behind the QITE-based approach is to restrict the basis of Pauli strings in ~(\ref{eq:pauli_expansion}) to those entering the UCCSD ansatz or a selection of them, typically using the Jordan-Wigner mapping \cite{jordan_1993}. Although this approach successfully converges, from a theoretical perspective it is incomplete, as general single Pauli strings do not encode particle number nor spin conservation.

We show in the supplementary material in \ref{subsec:alt_qite} that QITE can be reformulated in terms of a unitary ansatz $U=e^{\Delta\tau\hat{T}}$ that is the exponentiation of a general anti-hermitian operator $\hat{T}=-\hat{T}^\dagger$.

 $\hat{T}$ can be expanded in terms of a basis set:

\begin{equation}
\hat{T} = \sum_I a_I\hat{t}_I
\label{eq:ah_basis_set}
\end{equation}
\noindent
where every term in the basis is anti-hermitian, $\hat{t}_I^{\dagger}=-\hat{t}_I$ and coefficients $\{a_I\}$ are real-valued. 

In practice, if $\hat{t}_I$ is a Pauli string preceded by a $-i$ phase, $\hat{t_I} = -i \hat{\sigma}_{i_{1}}...\hat{\sigma}_{i_{D}}$, $\hat{\sigma}_i \in \{I,X,Y,Z\}$, the following formulation remains universal. In addition, $\hat{t}_I$ may be a particle number and spin-preserving n-tuple coupled-cluster excitation and de-excitation operator of the form:

\begin{equation}
\hat{t}_I = \hat{a}_{a_1}^{\dagger} \hat{a}_{a_2}^{\dagger} \cdots \hat{a}_{a_n}^{\dagger}\,
\hat{a}_{i_n} \cdots \hat{a}_{i_2} \hat{a}_{i_1} - h.c.
\label{eq:ah_ntuple}
\end{equation}

In terms of the general anti-hermitian basis in ~(\ref{eq:ah_basis_set}), the coefficients of QITE's linear system to second order are (see \ref{subsec:alt_qite}):

\begin{eqnarray}
    S_{I,J} = \bra{\Phi}\{\hat{t}_I,\hat{t}_J\}\ket{\Phi}
    \\
    b_I = \frac{1}{\sqrt{c}} (\bra{\Phi}[\hat{h}[m],\hat{t}_I]\ket{\Phi} - \frac{\Delta t}{2}\bra{\Phi}[\hat{h}[m]^2,\hat{t}_I]\ket{\Phi})
    \label{eq:new_formulas}
\end{eqnarray}

where $c$ is the same normalizing coefficient defined in ~(\ref{eq:linear_system2b}).

This approach presents the following advantages with respect to the Pauli string-based formulation:

\begin{enumerate}
  \item \textbf{Theoretical:} Particle number and spin are conserved along the evolution path. All operators enetering measurement are hermitian, and there is no need to resort to normal operators with complex eigenvalues.
  \item \textbf{Computational (classical):} The dimension of QITE's linear system is greatly reduced (roughly by a factor of 8).
  \item \textbf{Computational (quantum):} Using a singles and doubles ansatz and the Jordan-Wigner mapping \cite{jordan_1993}, all the hermitian operators entering ~(\ref{eq:new_formulas}) produce sets of Pauli strings that are mutually commuting. Applying suitable unitaries, as many as 64 Paulis can be measured simultaneously \cite{yen_2020}.
  \item \textbf{Circuit depth:} Single and double coupled-cluster excitation and de-excitation operators admit direct compilation in terms of Givens rotations \cite{yordanov_2020}, sparing as many as a factor 8 CNOT gates.
\end{enumerate}

\section{\label{sec:results}Results}

In this section, some demonstrations of the proposed MT-QITE algorithm are presented along with equivalent results using the previously reported QITE algorithm \cite{motta_2020}. The results correspond to noiseless emulation on classical hardware for four different physical systems, the XXZ Heisenberg model, the fermionic Hubbard model, the transverse-field Ising model and the $H_4$ chain. Numerical simulations are provided on 6 and 8 qubits, corresponding to an equal number of spins for the Ising and Heisenberg models, half that number of sites for the Hubbard model, and 8 spin-orbitals for the $H_4$ chain. The state fidelity with respect to the exact ground state is analyzed as a function of circuit depth and distinct Pauli measurements. Circuit depth is expressed as the number of necessary Pauli string rotations that implement the real-time evolution. In both cases MT-QITE is benchmarked against the optimum time-step size QITE original algorithm. There is a trade-off between accuracy and computational cost in all cases, and it is closely linked to the partition of the Hamiltonian employed.

Considerable classical pre-processing is required for the first three models in order to identify redundant Pauli strings according to $\mathbb{Z}_{2}$ symmetries in the Hamiltonian, as it is specified in \cite{sun_2021}. Once the pre-processing rules out incompatible Pauli strings, a choice of domain size $D=4$ for the 6-qubit problems and $D=5$ for the 8-qubit ones, reduces even further the number of relevant Pauli strings, and therefore the size of the linear system that must be measured and solved classically 

We benchmark MT-QITE with respect to the optimum time-step size QITE. Choosing a suitable time step size greatly impacts the fidelity, and unfortunately its value is model- and parameter-dependent. Numerical experiments show that a convenient strategy is to scan the $0.0 - 0.5$ time interval with 12 intermediate values, although the more values, the better fidelity at the expense of increasing the measurement count.

\subsection{Heisenberg model}

The XXZ Heisenberg Hamiltonian has a trivial mapping to qubits in terms of Pauli operators $\{\hat{\sigma}^x, \hat{\sigma}^y, \hat{\sigma}^z\}$:

\begin{equation}\label{eq:xxz_heis}
\hat{H} = \sum_{i=1}^{N-1}(\hat{\sigma}^x_{i}\hat{\sigma}^x_{i+1} +\hat{\sigma}^y_{i}\hat{\sigma}^y_{i+1}+J\hat{\sigma}^z_{i}\hat{\sigma}^z_{i+1})
\end{equation}

\noindent
for $N$ sites where $i$ denotes site (qubit) number, $J$ is the ZZ coupling constant and open boundary conditions are assumed.

The speed of convergence is strongly dependent on the choice of initial state. In order to have comparable trends, a batch of 25 symmetry-compatible initial states is prepared for 8-qubit simulations, and each initial state is used to run both MT-QITE and the original QITE. A realistic procedure to produce initial states scattered across the invariant subspace is the following: we start always with the trivial state $\ket{000...0}$ and apply an X gate followed by an H gate on each qubit, each with a given probability (apply / do not apply). The resulting product state is then evolved using exact ITE according to the equally weighted sum of all Pauli strings in the stabilizer of the Hamiltonian, i.e. according to an auxiliary operator whose ground state is highly degenerate and spans the whole invariant subspace.

The simulation results are plotted in FIG. \ref{fig:fig_heatm} for 6 qubits and a selection of ZZ coupling strengths (J), and in FIG. \ref{fig:fig_main} for 8 qubits and the challenging phase transition (J = 1) along with standard deviations of fidelity. Clearly, MT-QITE improves ground state fidelity with respect to QITE in one order of magnitude after 10 Trotter steps, attaining exact fidelity up to the fifth decimal position, and using a circuit consisting of $\sim$ 1200 Pauli string rotations. In addition, MT-QITE involves a measurement budget of the order of $10^5$ distinct Pauli strings, one order of magnitude below QITE's requirements.

\begin{figure*}
    \centering
    \begin{subfigure}{0.4\textwidth}
        \centering
        \includegraphics[width=\textwidth]{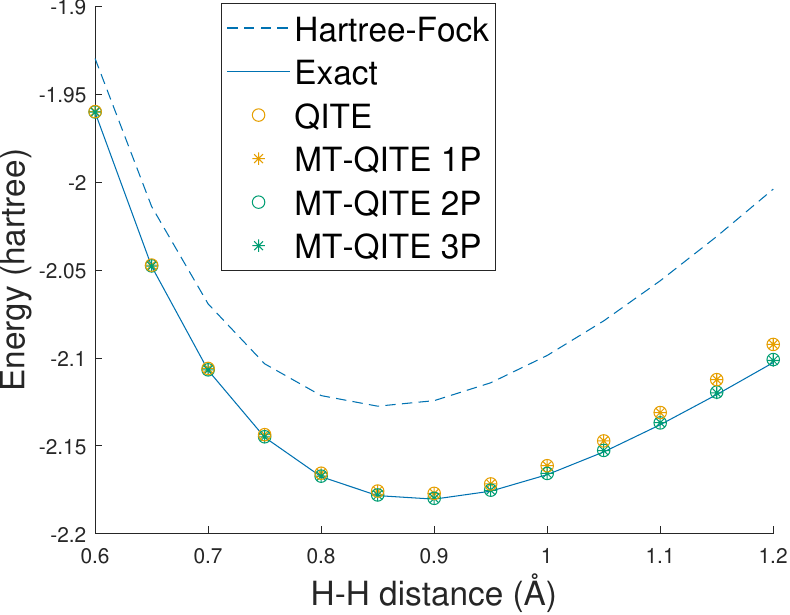}
        \caption{$H_4$ binding energy curve}
        \label{fig:first}
    \end{subfigure}
    \begin{subfigure}{0.4\textwidth}
        \centering
        \includegraphics[width=\textwidth]{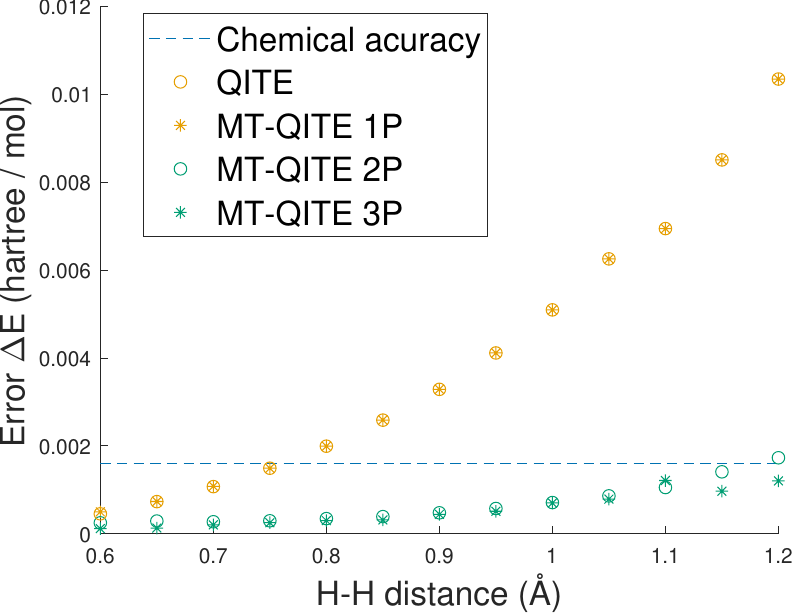}
        \caption{Absolute error in energy}
        \label{fig:second}
    \end{subfigure}
    \caption{Energy and its absolute error for the $H_4$ chain in function of interatomic distance. Simulations have been run using the pool of anti-hermitian operators given by the UCCGSD ansatz. For stretching geometries, where more correlation is present, QITE and MT-QITE without partitioning (1P) fail, whereas MT-QITE on a three-term partition (3P) converges to chemical accuracy after 6 Trotter steps for the whole span of interatomic distances, and MT-QITE on a two-term partition (2P) follows closely.}
    \label{fig:chem}
\end{figure*}

\subsection{Fermionic Hubbard model}

The fermionic Hubbard model obeys the following Hamiltonian:

\begin{equation}\label{eq:ferm_hubb}
\hat{H} = -\sum_{<i,j>,\sigma}(\hat{a}^\dagger_{i\sigma}\hat{a}_{j\sigma} +\hat{a}^\dagger_{j\sigma}\hat{a}_{i\sigma})+U\sum_{i}\hat{n}_{i\uparrow}\hat{n}_{i\downarrow}
\end{equation}   

\noindent
where $i, j$ denote sites, there are two fermionic orbitals with opposite spin orientation $\sigma$ per site, and the summation over $<i,j>$ corresponds to adjacent sites, again assuming open boundary conditions. $\hat{a}^\dagger$ and $\hat{a}$ are the fermionic creation and annihilation operators respectively, and $\hat{n}$ is the number operator. In this case, mapping to qubits is non-trivial and the Jordan-Wigner transformation \cite{jordan_1993} is used to preserve particle statistics. 

In FIG. \ref{fig:fig_heatm} the performance of MT-QITE vs QITE is plotted for a selection of repulsion constant values ($U$) and a common initial state with a well-defined number of particles. Accuracy is harder to achieve for high repulsive strengths, but again, MT-QITE outperforms QITE after 10 Trotter in both fidelity and number of measurements.
\\

\subsection{Transverse-field Ising model}

The Ising Hamiltonian also has the following trivial mapping to qubits in terms of Pauli operators $\{\hat{\sigma}^x, \hat{\sigma}^y, \hat{\sigma}^z\}$:

\begin{equation}\label{eq:ising_model}
\hat{H} = -\sum_{i=1}^{N-1}(\hat{\sigma}^z_{i}\hat{\sigma}^z_{i+1}) +(h/J)\sum_{i=1}^{N}\hat{\sigma}^x_{i}
\end{equation}

\noindent
for $N$ sites where $i$ denotes site (qubit) number, $h/J$ is the transverse field with respect to the ZZ coupling, and open boundary conditions are assumed.

Simulations are displayed for 6 qubits varying the strength of the transverse field in FIG. \ref{fig:fig_heatm}, and changing the configuration of partitions in FIG. \ref{fig:fig_main}. In the latter case the hamiltonian is split in three terms, with two of them equivalent under site inversion symmetry. Simulations are run for QITE and MT-QITE, both grouping the equivalent terms (two-term partition, 2P), and ungrouping them and taking advantage of symmetry (three-term partition, 3P). MT-QITE outperforms QITE in two orders of magnitude in fidelity, and one order of magnitude in computational cost. In addition, it becomes clear that reduction by symmetry reduces further the measurement count.

\subsection{$H_4$ chain}

The $H_4$ chain provides a compact, yet non-trivial, benchmark for evaluating the performance of quantum chemistry algorithms. Consisting of four hydrogen atoms arranged linearly, the system exhibits strong electron correlation in stretched geometries. Using standard package PySCF \cite{sun_2018} and the STO-3G basis \cite{hehre_1969}, one- and two-body integrals are calculated, and applying the Jordan-Wigner mapping, the second-quantized Hamiltonian in ~(\ref{eq:2ndq_ham}) is implemented on 8 qubits for a varying interatomic distance ranging from 0.60 to 1.20 \AA. We use a unitary coupled-cluster ansatz of generalized singles and doubles (UCCGSD) \cite{anand_2022} and we benchmark MT-QITE against QITE using the same pool of antihermitian excitation and de-excitation operators as described in section \ref{subsec:mtqitechem}. In FIG. \ref{fig:chem}, the binding energy curve is plotted in function of interatomic distance for QITE and MT-QITE, without partitioning and using a two-term and a three-term partition. These Hamiltonian groupings have been obtained as those that maximize the commuting structure of terms in the Hamiltonian with respect to terms in the ansatz. After 6 Trotter steps, we observe clearly how QITE and MT-QITE without partitioning struggle to converge with increasing distance, while MT-QITE on the three-term partition reaches chemical accuracy even for challenging interatomic separations. This is remarkable, as partition terms are agnostic of each other, and no measurement overhead is introduced to calculate QITE linear systems. Counterintuitive as it may seem, in this case, there is an advantage in partitioning with a Hamiltonian comprising non-local interactions.

\section{\label{sec:conclusion}Conclusions and outlook}

We have presented the Multiple-Time Quantum Imaginary Time Evolution algorithm (MT-QITE) and shown that using multiple imaginary times improves both ground state fidelity and computational cost substantially with respect to the previously reported QITE algorithm. Unlike other developments using ITE on quantum hardware available in the literature, QITE and MT-QITE are approximation-free and non-probabilistic.

We have shown for four foundational models how MT-QITE outperforms QITE in ground state fidelity by one or two orders of magnitude after 10 Trotter steps, while requiring around 10 times less measurements. This makes MT-QITE the ideal candidate when higher fidelity at constant circuit depth is required, avoiding approximations and probabilistic subroutines.

We have introduced an alternative formulation of the linear-system-based QITE routine especially well suited for quantum chemistry Hamiltonians. We have proved that it successfully captures correlation in the $H_4$ chain and that even for this non-local Hamiltonian, there is an advantage in partitioning and using several imaginary times. Moreover, this opens the way to parallelizing real-world Hamiltonians with thousands of terms, as MT-QITE treats partitial Hamiltonians completeley agnostic of each other.

More work is needed to quantify the trade-off between fidelity, circuit depth and measurement count and its dependence on the choice of domain and partition of the Hamiltonian, as well as the resolution of the linear systems of equations with pseudoinverse matrices. In particular exploiting 3- or more-term partitions needs further investigation, as the bottleneck of the algorithm is the linear system size, and this is reduced by partitioning. The variational optimization introduced by MT-QITE should not be a problem for multiple-term partitions, as the position of the minimum tends to the smaller time step sizes as Trotter steps proceed, and procedures like gradient descent can be considered.

%
%
\subsection*{Acknowledgements}

This work has been supported by: MG s financial support from the Knut and Alice Wallenberg Foundation through the Wallenberg Centre for Quantum Technology (WACQT). EvN acknowledges fruitful discussions with Stefano Polla and Jordi Tura, and that this work was supported by the Dutch National Growth Fund (NGF), as part of the Quantum Delta NL programme.

\subsection*{Code availability}

The code used to obtain all the results in the present paper is available in Matlab in the following GitHub repository: https://github.com/jdelcasti26/fermi-lab .

\section{\label{sec:supplement}Supplementary material}

\begin{figure*}[t]
    \centering
    \includegraphics[width=0.8\textwidth]{fig_alg}
    \caption{\label{fig:alg}Comparison between the previously reported QITE algorithm and MT-QITE, illustrating the wavefunctions needed at each step for a Hamiltonian split in a two-term partition, $h[1]$ and $h[2]$, two Trotter steps, and $L$ different time step sizes. Kets in white boxes represent the different reference wavefunctions that must be prepared repeatedly and subjected to Pauli measurements in order to determine the real time evolution (RTE) that approximates the imaginary time evolution towards the ground state. Blue boxes indicate that new measurements are needed, and gray boxes represent classical processing of the available measurements taking into account the desired $\Delta t$. Clearly, the measurement budget of QITE grows linearly with the number of different time step sizes $L$, whereas for MT-QITE it does not. In addition, using common reference states allows either parallelization of the different terms in the partition, or it may result in a reduced measurement count if the Hamiltonian terms share a common support. Shortcuts by symmetry are also possible when a common reference state is used. Notice that times evolve vertically on the graph as Trotter steps proceed.}
\end{figure*}

In this section we provide a graphical representation of the main differences between the QITE and the MT-QITE algorithms in FIG. \ref{fig:alg}, we derive the QITE formulas ~(\ref{eq:new_formulas}) in terms of a general anti-hemitian operator basis in subsection \ref{subsec:alt_qite} and we compare algorithm performance between MT-QITE and related work in subsection \ref{subsec:similar_work}.

\subsection{Alternative derivation of the QITE formulas}
\label{subsec:alt_qite}

We include here the explicit derivation of the QITE formulas in function of a basis of arbitrary anti-hermitian operators and show how the original formulas can be reinterpreted without the need of measuring normal operators with potentially imaginary expectation values.

The goal is to approximate the normalized ITE evolution using a unitary operator $U$ such as:

\begin{equation}
\ket{\Phi'} = \frac{e^{-\Delta\tau{\hat{h}[m]}}\ket{\Phi}}{\lVert e^{-\Delta\tau{\hat{h}[m]}}\ket{\Phi}\rVert} \approx U\ket{\Phi} = e^{-i\Delta\tau{\hat{A}[m]}}\ket{\Phi}
\label{eq:unit_approx2}
\end{equation}

In what follows, we assume a single hermitian Hamiltonian term $\hat{h}[m]$ and we drop the index $m$. We define the square of the normalization constant as:

\begin{equation}
c = \bra{\Phi}e^{-2\Delta\tau{\hat{h}}}\ket{\Phi}
\label{eq:norm_const}
\end{equation}

Its second order approximation being:

\begin{equation}
c \approx 1 - 2\Delta\tau \bra{\Phi}\hat{h}\ket{\Phi} + {\Delta\tau}^2 \bra{\Phi}\hat{h}^2\ket{\Phi}
\label{eq:norm_const1}
\end{equation}

Without loss of generality, we consider a general unitary ansatz $U=e^{\Delta\tau\hat{T}}$, where $\hat{T}$ is anti-hermitian and can be expanded in terms of an anti-hermitain basis as in ~(\ref{eq:ah_basis_set}), $\hat{T} = \sum_J a_J\hat{t}_J$ with real coefficients $\{a_J\}$. The goal is to produce a unitary evolution, to first order

\begin{equation}
\ket{\Delta} = (\sum_J a_J\hat{t}_J)\ket{\Phi}
\end{equation}

that is as close as possible to the ITE evolution

\begin{equation}
\ket{\Delta_0} = \frac{\ket{\Phi'}-\ket{\Phi}}{\Delta\tau}
\end{equation}

We define:

\begin{equation}
    N_0(a_1,\dots,a_I,\dots) = {\lVert \ket{\Delta - \Delta_0} \rVert}^2
\end{equation}

and thus the objective is to minimize $N_0$ with the conditions:

\begin{equation}
\frac{\partial N_0}{\partial a_I} = 0
\end{equation}

Expanding the definition of $\ket{\Delta_0}$ to second order we obtain:

\begin{equation}
    \ket{\Delta_0} = (\frac{c^{-1/2}-1}{\Delta\tau}\hat{1}-c^{-1/2}(\hat{h}-\frac{\Delta\tau}{2}\hat{h^2}))\ket{\Phi}
\end{equation}

For clarity, we define the hermitian operator
\begin{equation}
    \hat{\mathbf{H}} = \frac{c^{-1/2}-1}{\Delta\tau}\hat{1}-c^{-1/2}(\hat{h}-\frac{\Delta\tau}{2}\hat{h^2})
\end{equation}

And thus $N_0$ can be written as:

\begin{eqnarray}
    N_0 = \bra{\Delta_0-\Delta}\ket{\Delta_0-\Delta}
    \label{eq:N_01}
    \\
    N_0 = \bra{\Phi}(\hat{\mathbf{H}}+\sum_J a_J\hat{t}_J)(\hat{\mathbf{H}}-\sum_J a_J\hat{t}_J)\ket{\Phi}
    \label{eq:N_02}
    \\
    N_0 = \bra{\Phi}\hat{\mathbf{H}}^2\ket{\Phi}+ \bra{\Phi}\sum_J a_J[\hat{t}_J,\mathbf{H}]\ket{\Phi}
    \\
    \nonumber
    -\bra{\Phi}\sum_J a_J\hat{t}_J\sum_K a_K\hat{t}_K\ket{\Phi}
\end{eqnarray}

\begin{figure*}[t]
    \centering 
    \begin{subfigure}{0.32\textwidth}
        \centering
        \includegraphics[width=\linewidth]{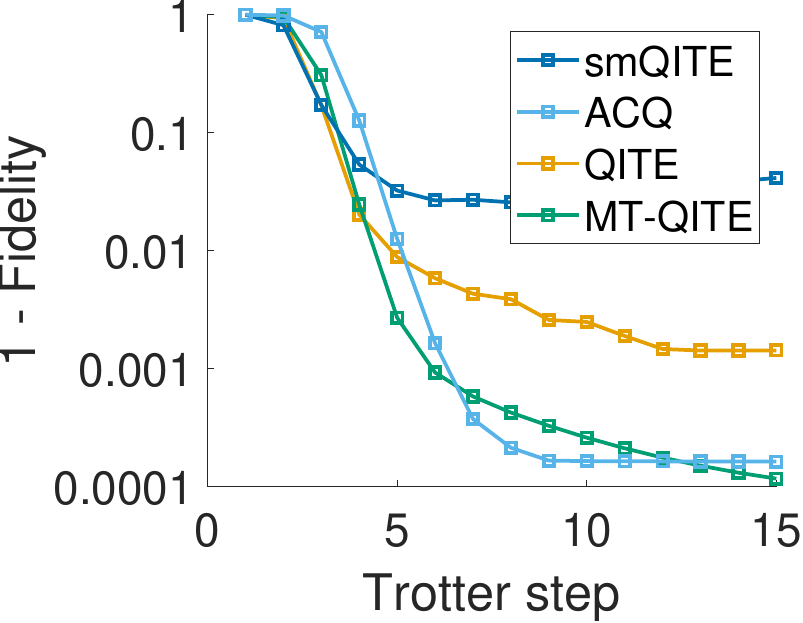}
        \label{fig:1a}
    \end{subfigure}\hfill
    \begin{subfigure}{0.32\textwidth}
        \centering
        \includegraphics[width=\linewidth]{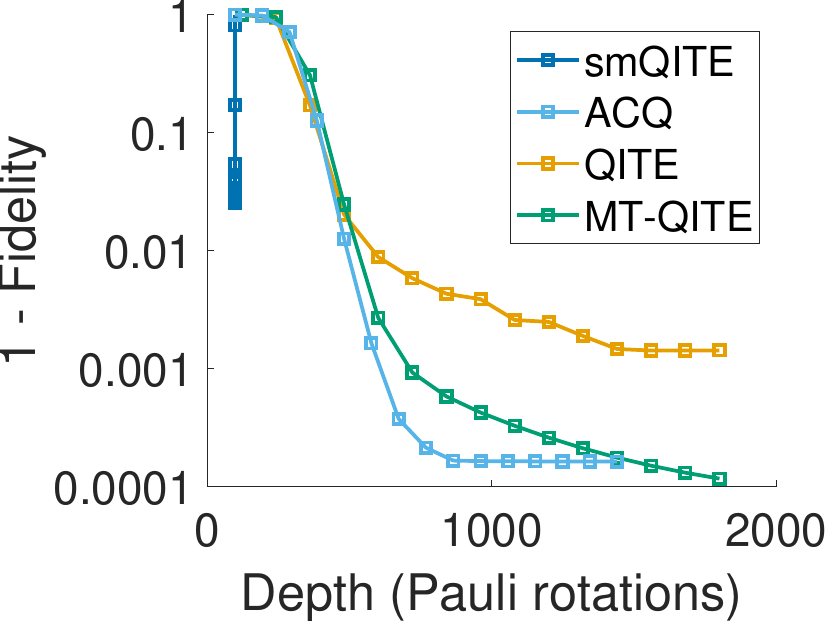}
        \label{fig:1b}
    \end{subfigure}\hfill
    \begin{subfigure}{0.32\textwidth}
        \centering
        \includegraphics[width=\linewidth]{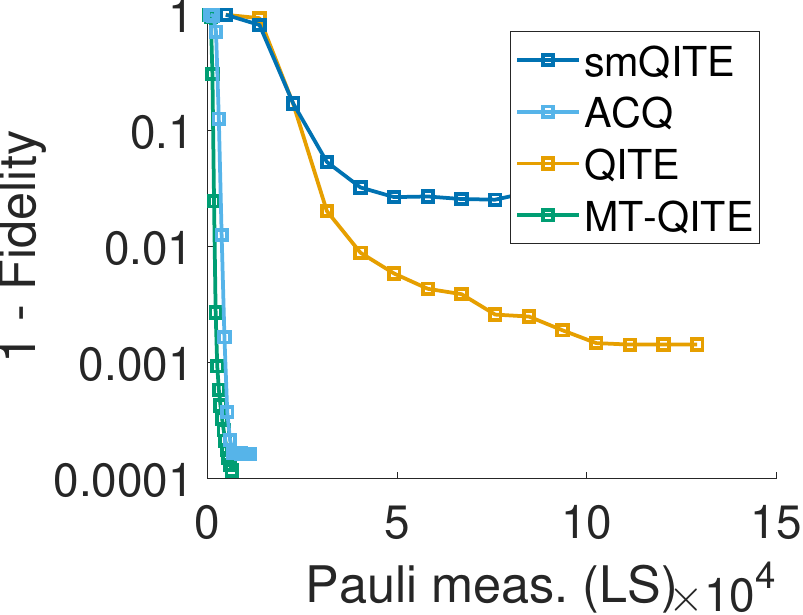}
        \label{fig:1c}
    \end{subfigure}
    \caption{Algorithm comparison for the 6-spin Heisenberg model with open boundaries using a two-term symmetric partition. Ground state fidelity is monitored vs. Trotter step, circuit depth in number of Pauli string rotations and distinct Pauli measurements to build the QITE linear system. Both ACQ and MT-QITE (with native time-step optimization) outperform optimal-time-step QITE and smQITE in fidelity and measurement budget. ACQ is provide shallower circuits than MT-QITE at the expense of plateuing before. In terms of measurement count, MT-QITE outperforms ACQ by a factor 2, and crucially, MT-QITE permits running the different partition terms in parallel.}
    \label{fig:row1}
\end{figure*}
\begin{figure*}[t]
    \centering
    \begin{subfigure}{0.32\textwidth}
        \centering
        \includegraphics[width=\linewidth]{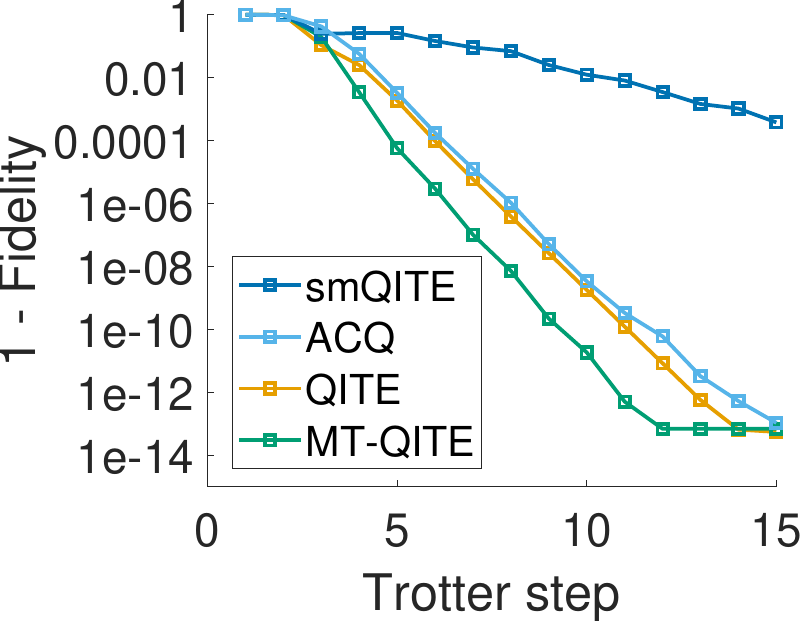}
        \label{fig:2a}
    \end{subfigure}\hfill
    \begin{subfigure}{0.32\textwidth}
        \centering
        \includegraphics[width=\linewidth]{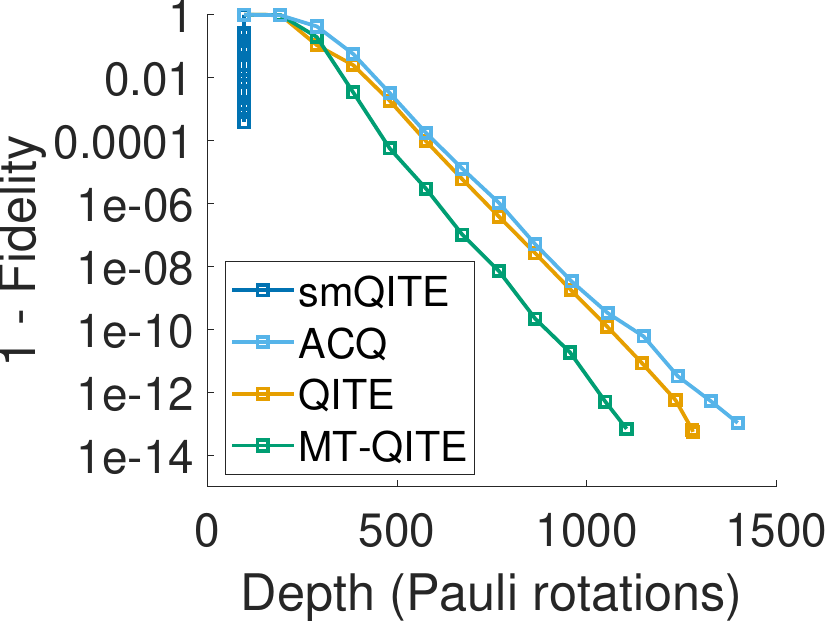}
        \label{fig:2b}
    \end{subfigure}\hfill
    \begin{subfigure}{0.32\textwidth}
        \centering
        \includegraphics[width=\linewidth]{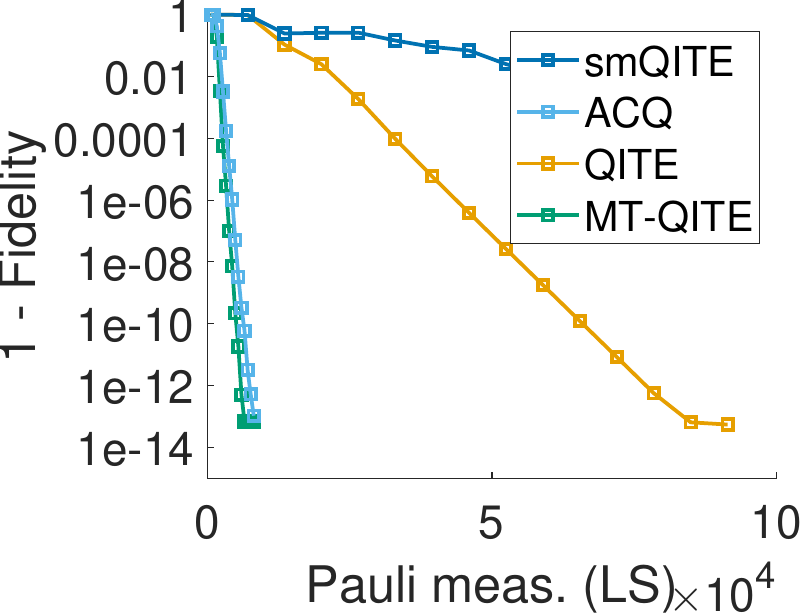}
        \label{fig:2c}
    \end{subfigure}
    \caption{Algorithm comparison for the open-boundary 6-spin Heisenberg model without partitioning and thus free of Trotter error. In this case, all the algorithms converge to higher fidelities than the paritioned counterparts, but ACQ and MT-QITE are clearly superior to QITE in terms of computational cost. MT-QITE reaches the same accuracy plateau as ACQ around $10^{-13}$ but does so three Trotter steps in advance and therefore sparing about 300 Pauli rotations.}
    \label{fig:row2}
\end{figure*}

Taking partial derivatives with respect to $a_I$, we obtain the system of linear equations:

\begin{eqnarray}
    0 = \bra{\Phi}[\hat{t}_I,\mathbf{H}]\ket{\Phi}
    \\
    \nonumber
    -\bra{\Phi}\hat{t}_I\sum_J a_J\hat{t}_J+(\sum_J a_J\hat{t}_J)\hat{t}_I\ket{\Phi}
    \\
    \bra{\Phi}[\hat{t}_I,\mathbf{H}]\ket{\Phi} = \sum_J a_J \bra{\Phi}\{\hat{t}_I,\hat{t}_J\}\ket{\Phi}
\end{eqnarray}

The commutator involving $\mathbf{H}$ can be simplified, as $\hat{1}$ commutes with any operator, so we have QITE's linear system of equations $\textbf{S}\textbf{a}[m]=\textbf{b}$, where:

\begin{eqnarray}
    b_I = \frac{1}{\sqrt{c}} (\bra{\Phi}[\hat{h},\hat{t}_I]\ket{\Phi} - \frac{\Delta\tau}{2}\bra{\Phi}[\hat{h}^2,\hat{t}_I]\ket{\Phi})
    \\
    S_{I,J} = \bra{\Phi}\{\hat{t}_I,\hat{t}_J\}\ket{\Phi}
\end{eqnarray}

Note that unlike the original formulation, $[\hat{h},\hat{t}_I]$, $[\hat{h}^2,\hat{t}_I]$ and $\{\hat{t}_I,\hat{t}_J\}$ are all hermitian. In the case of using the Pauli basis, every $\hat{t}_I$ corresponds to a Pauli string preceded by a $-i$ phase, i.e. $\hat{t}_I = -i\hat{\sigma}_I$. As it turns out, if two Pauli strings commute, their product carries a $\pm$ phase, commutators are zero and the anti-commutator introduces a factor $2$ (see original formulas below). If two Pauli strings anti-commute, their product produces a $\pm i$ phase, the anti-commutator vanishes and the commutator carries a two factor. In every case, in this alternative formulation, every measurement involves a Pauli string with a real phase. 

\begin{eqnarray}
c = 1 - 2\Delta\tau\bra{\Phi}\hat{h}[m]\ket{\Phi} + 2\Delta\tau^2\bra{\Phi}\hat{h}[m]^2\ket{\Phi}
\label{eq:linear_system2bprime}
\\
S_{I,J} = 2 Re \bra{\Phi}\sigma^{\dagger}_I\sigma_J\ket{\Phi}
\label{eq:linear_system3bprime}
\\
b_{I} = 2 Im \frac{1}{\sqrt{c}} \bra{\Phi}{\sigma^{\dagger}_I\hat{h}[m] - \frac{\Delta\tau}{2}\sigma^{\dagger}_I\hat{h}[m]^2}\ket{\Phi}
\label{eq:linear_sys2bprime}
\end{eqnarray}

\subsection{Comparison with similar work}
\label{subsec:similar_work}
We provide here (FIG. \ref{fig:row1} and FIG. \ref{fig:row2}) comparative results of the MT-QITE algorithm with respect to similar work. In particular, the Step-Merged QITE algorithm (smQITE) as described in \cite{gomes_2020}, the more recent Adaptive-Time-Compressed QITE (ACQ) presented in \cite{melendez_2025} and the original QITE \cite{motta_2020}. Both ACQ and MT-QITE are deterministic as QITE, and in addition they allow a reduced measurement budget along with improved fidelity while providing time step optimization natively. We provide simulations on 6 qubits for the Heisenberg model for a symmetric two-term partition and without partitioning, benchmarking fidelity vs Trotter step, circuit depth and required measurements to solve the QITE linear systems, the bottleneck of all these algorithms when it comes to scaling. ACQ and MT-QITE outperform smQITE and QITE in fidelity and measurement count, whereas smQITE provides extraordinarily shallow circuits at the expense of compromising fidelity. ACQ and MT-QITE have similar performance and present slight advantages compared to each other depending on the setting, but crucially, MT-QITE is parallelizable.

\bibliography{references}

\end{document}